\documentclass[11pt,tightenlines,eqsecnum,floats,aps,nofootinbib,prd,showpacs]{revtex4}
\usepackage{hyperref}
\usepackage{amsmath,amssymb,amsfonts,amsthm,amscd}
\usepackage{graphicx}
\usepackage{enumerate}
\usepackage{colordvi}
\usepackage{units}
\usepackage{epsfig}
\usepackage{natbib}
\usepackage{enumerate}
\usepackage{colordvi}
\usepackage{multirow}
\usepackage{afterpage}
\usepackage{pdflscape}
\usepackage{subfig}
\usepackage[utf8]{inputenc}

\def\be{\begin{equation}}
\def\ee{\end{equation}}
\def\ba{\begin{eqnarray}}
\def\ea{\end{eqnarray}}

\def\State{\mathfrak{s}}
\def\Lie{\mathfrak{L}}
\def\L{\mathcal{L}}
\def\H{\mathcal{H}}
\def\D{\mathbf{D}}
\def\F{\mathbf{F}}
\def\G{\mathbf{G}}
\def\V{\mathbf{V}}

\def\X{\mathbf{X}}
\def\M{\mathcal{M}}

\begin{document}

\title{Dynamical Similarity}

\author{David Sloan}
\email{david.sloan@physics.ox.ac.uk}
\affiliation{Beecroft Institute of Particle Astrophysics and Cosmology, Department of Physics,
University of Oxford, Denys Wilkinson Building, 1 Keble Road, Oxford, OX1 3RH, UK}

\begin{abstract}
\noindent We examine ``dynamical similarities" in the Lagrangian framework. These are symmetries of an intrinsically determined physical system under which observables remain unaffected, but the extraneous information is changed. We establish three central results in this context: i) Given a system with such a symmetry there exists a system of invariants which form a subalgebra of phase space, whose evolution is autonomous; ii) this subalgebra of autonomous observables evolves as a contact system, in which the friction-like term describes evolution along the direction of similarity; iii) the contact Hamiltonian and one-form are invariants, and reproduce the dynamics of the invariants. As the subalgebra of invariants is smaller than phase space, dynamics is determined only in terms of this smaller space. We show how to obtain the contact system from the symplectic system, and the embedding which inverts the process. These results are then illustrated in the case of homogeneous Lagrangians, including flat cosmologies minimally coupled to matter; the n-body problem and homogeneous, anisotropic cosmology.

\end{abstract}

\maketitle

\section{Introduction}
\noindent
``Dynamical Similarity" is a term that has come to mean the intrinsic indistinguishability of solutions in relational theories \cite{Mercati:2014ama,Gomes:2012hq}. In a framework in which the measuring apparatus (rods and clocks) has to be established in the same system as the observed physics, there is often a redundancy that lies in assigning properties to the rods and clocks that are chosen from within the system. A real world example is that we declare by fiat that the meter stick in Paris has a fixed length which is unchanging, and that the oscillation of a caesium atom has a fixed period. This is often done for reasons of convenience. In our everyday physics the ratio of the diameter of the Earth to that of the meter stick is unchanging, as is the ratio of a the radius of a hydrogen atom to either of these. Therefore it appears natural to choose that all these have trivial evolutions in our models, and use any one of these elements to define an external parameter against which we evaluate subsystems. A similar property holds for caesium oscillations, pendulum clocks and the rotation of the Earth about its axis, and orbit around the sun. However, it is particularly apparent in cosmology, for example, that this is not a well motivated choice everywhere: If we adopt such a notion of length and time, in the past such systems have no fixed rods or clocks according to this definition, since the gravitational force overwhelms the other forces keeping these objects fixed. Our goal is to describe the reduced space of systems modulo this freedom to make rescalings, as this should not affect the physical observables, but map between descriptions according to different choices of rod or clock \cite{Wetterich:2014zta}. In the current paper we will restrict ourselves to working only with vector fields on phase space, leaving the couplings fixed. This is done so that we can establish the mathematical structure of the theory in a simple setting. In future work we will expand this to include rescaling also the couplings in a manner such that the intrinisic physics is unaltered. 

Since we are interested in transformations which preserve the dynamics of our theory, these transformations must preserve the form of conserved quantities. In particular, level surfaces of the Hamiltonian should be mapped onto level surfaces, hence $\H$ can be rescaled but not deformed.  From the intrinsic perspective, this is to be expected, as time is not a direct observable of our system, but rather must be inferred from observations of subsystems which we choose as clocks. A transformation which both halves a velocity, and the rate of the clock with which progress is measured creates a dynamical similarity between the two solutions - these have the same evolution, just with a rescaled time coordinate. A helpful model to keep in mind (which we will discuss at greater length later) is that of orbits under a central potential sourcing an inverse square law. For each orbit with semi-latus rectum $l$, there exists a similar orbit with semi-latus rectum $\alpha l$. Observers on each planet would term their orbit times to be one `year', locally defined and related by a change of lapse. The energies and angular momenta of the orbits will rescale in a similar manner to the orbital period. This retains symmetry at the level of the action since the initial and final states $\State_1$ and $\State_2$ which limit the integral should be functions of intrinsic observables - the interval in $t$ over which the Lagrangian is to be integrated will vary. Below we show explicitly how this comes to happen.

In the following section we will give a precise definition of the action of dynamical similarity on phase space. Then, in section \ref{Contact} we show that the dynamics of the reduced space consisting of the algebra of invariants 
of the dynamical similarity behaves as a contact space. Here we will establish the equivalence between the two frameworks, and show that the dissipative description in terms of a contact manifold matches with the idea of an ``arrow of time" in the manner introduced by Barbour et al. This is generalized beyond the informative examples of the Newtonian n-body problem to any Hamiltonian theory which exhibits a dynamical similarity. In section \ref{Homogeneous} we give an example class of Lagrangian systems, those homogeneous in a configuration variable, which exhibit this similarity, and show explicitly the construction of the contact system and the resulting dissipative dynamics. A particular illustrative example shows that for a broad class of cosmological models, the arrow of time introduced points in the direction of the expansion of space, and the ``Janus points" of qualitative similarity along physical trajectory correspond to points of bounce or recollapse. In section \ref{NBody} we show that the general n-body system with potential homogeneous in separation of particles (but of arbitrary power) is dynamically similar under rescaling. From the contact description of this we form the shape space, and show that there is a natural extension of the results of Barbour et al to generic potentials. The recent results of the continuation of homogeneous cosmology beyond the big bang singularity \cite{Koslowski:2016hds} are placed within this context in section \ref{Cosmology}. Here we do not reproduce the singularity result, but rather focus on how dynamical similarity reveals the existence of the autonomous system which remains well defined. Finally in section \ref{Discussion} we will remark upon the significance of the results and future directions. 

\section{Generating Dynamical Similarities} \label{Generating}

\noindent  We will begin our analysis with an action, the minimization of which will provide our equations of motion \footnote{The author is grateful to Jos\'e Cari\~nena a number of useful remarks and clarifications on an earlier version of this section}. This will consist of the integral of a Lagrangian one-form over a space of configurations $q$ and their velocities $\dot{q}$.
\be S = \int_{\State_1}^{\State_2} \L  \ee
 Throughout this discussion, although we will use time, $t$, as a parametrization of a solution curve, we will only consider (a subset of) the $q$ and $\dot{q}$ to be physically observable. Our motivation is that we want to describe intrinsic, relational physics. As such observations are not made directly of time, but of a clock variable which represents time. In a similar fashion, we will choose to consider any dimensionful variable not to be directly observable, but only to be observable in dimensionless ratios with other variables. Given a configuration space $Q$, the phase space, $\Gamma=T^*Q$, is the cotangent bundle over the configuration space. A symplectomorphism is a diffeomorphism $f: \Gamma \rightarrow \Gamma$ whose action is to preserve the symplectic structure under pull-back: $f^* \omega=\omega$. Consider a vector field $\V$ consisting of a flow $\phi_t$ generating such diffeomorphisms; its action is to Lie drag the symplectic structure: $\Lie_\V \omega=0$. Such a vector field is called `locally Hamiltonian'. In previous work some of the basic motivation behind dynamical similarity was examined as an extension of this to a a non-strictly canonical transformation \cite{Carinena:2013zpa,Carinena:2014bda} $f:\Gamma \rightarrow \Gamma$ under which $f^*\omega = a \omega$ for $a \in \mathcal{R}$, $a$ is known as the valence of the transformation. The vector field generating this on phase space is also referred to as a `Liouville vector field' in the literature \cite{GEIGES19971193} \footnote{In earlier works this was referred to as a 'scaled symplectomorphism'}. 
 
We will be interested in projecting down under the action of a continuous symmetry on phase space, we will be considering symmetries which are generated by flows on phase space. The Lie derivative along such vector fields is to propagate the symplectic structure rescaled; $\Lie_\G \omega = \lambda \omega$. This is a non-strictly canonical transformation. Since $\G$ is a vector field, it is linear, and thus without loss of generality we can fix $\lambda=1$. This is a quite general construction; consider the flow $\phi_t$ such that $\phi_t^* \omega = r(t)\omega$, for some strictly positive function $r(t)$. Reparametrizing the flow we can enforce $r(t)=\exp(t)$ and hence arrive at the result. It is clear from this definition that non-strictly canonical transformations, (like their strict counterparts) comprise a group under addition. A direct evaluation shows the Lie bracket of any two such transformations of the same valence is locally Hamiltonian. Due to Cartan's identity
\be \Lie_X = d \circ \iota_X + \iota_X \circ d \ee
and the fact that the symplectic two form is exact (being the exterior derivative of the symplectic potential, $\omega = d\theta$) evaluating the action of the symplectic vector field is simply
\be \Lie_\G \omega = d \circ \iota_\G \omega \ee
and hence we know that $\G$ is a non-strictly canonical transformation if $\iota_\G \omega = \theta + d\xi $ for some exact form $d\xi$, and hence we note that 
\be \Lie_\G \theta = d \circ \iota_\G \theta + \iota_\G \circ d\theta = d \circ \iota_\G \theta + \iota_\G \omega = \theta +d(\xi+ \iota_\G \theta) \ee
In this we have made the derivative term, $d(\xi+\iota_\G \theta)$ explicit to highlight that this only contributes a boundary term to the Lagrangian under Lie dragging by our symmetry generator. Equivalently, this can be surmised from the commutativity of the Lie and exterior derivatives, which together with $\G$ being locally Hamiltonian implies $\Lie_\G \theta=\theta + \chi$ for some exact form $\chi$. In practical circumstances it is often simpler to verify that the difference between $\iota_\G \omega$ and $\theta$ is a closed form - i.e. $d \circ \iota_\G \omega= \omega$. 

Physics in our system is determined by the Hamiltonian function $\H$ which generates time evolution. The Hamiltonian flow is determined by $d\H = \iota_{\X_\H} \omega$ (note that this is unique up to a constant due to the fibre-wise invertibility of $\omega$, giving the Poisson structure). Consider now a diffeomorphism $g: \Gamma \rightarrow \Gamma$ whose inverse $f$ both rescales the symplectic structure and the Hamiltonian $\H$ alike, and whose flow consists of a vector field $\G$; 
\be f^*\omega = \lambda \omega \quad f^* \H = a \H \ee
As above, the vector field $\G$ is non-strictly canonical, and acts to rescale the Hamiltonian. We shall call such vector fields the generators of dynamical similarity for reasons which will become obvious. Following the normalization conventions above;
\be \G \H = \Lambda \H \quad \Lie_\G \omega = \omega \ee
Such a transformation acts only to rescale the Hamiltonian flow:
\be \iota([\X_\H,\G])\omega = (\iota_{\X_\H} \Lie_\G - \Lie_\G \iota_{\X_\H})\omega = (\Lambda-1)dH \label{SymplecticScale} \ee
and hence from the non-degeneracy of $\omega$;
\be [ \X_\H, \G] = (\Lambda-1) \X_\H \ee
It is important to note here that since $\G$ only acts on $\H$ through its action on phase space variables, the transformation is only valid in the case where $\H$ is a function of phase space variables alone. This follows a light generalization of an argument presented in \cite{Carinena:2013zpa} and \cite{2012JPhAC}. Note that this will necessarily change the non-zero energy of the system. This is not unexpected, indeed transforming reference frames between solutions (e.g. working in center of mass coordinates) alters the energy of the system by a constant. The role of the Hamiltonian (i.e. the generator of time evolution) $\H$ in this construction is in fact unimportant; we could have chosen the generator of any Hamiltonian vector field (generating a conserved quantity of the system). For a Hamiltonian vector field, there is a conserved quantity $C$, defined $dC=\iota_\V \omega$. These commute with the Hamiltonian flow (as they are constant): $[\X_\H,\V]=0$. From Jacobi's identity;
\be [\G, [\X_\H,\V]] + [\X_\H,[\V,\G]] + [\V, [\G, \X_\H]] = 0 \ee
we see that the first and third terms are identically zero and hence $[\V,\G]$ is a Hamiltonian vector field. Hence the action of $\G$ is to map conserved quantities onto conserved quantities. 

Given an invariant $\State$ of $\G$ we are able to determine the action of $\G$ on the one-form $dt$:
\be \Lie_\G dt = \Lie_\G \left(\frac{d\State}{\iota_{d\State}\X_\H }\right) = \frac{(1-\Lambda) \iota_{d\State} \X_\H}{(\iota_{d\State} \X_\H)^2} = (1-\Lambda) \frac{d\State}{\iota_{d\State} \X_\H} = (1-\Lambda) dt \ee
Thus we see that the transformation of the one-form $\H dt$ is simple: $\Lie_\G \H dt = \H dt$ - it scales in exactly the same way as the symplectic potential. Therefore we can present a powerful result: Given a generator of dynamical similarity, $\G$ the Lagrangian $\L$ is rescaled by the the Lie derivative along $\G$. This arises directly as an application of the above and equation (\ref{SymplecticScale}), up to boundary terms we see:
\be \Lie_\G \L = \Lie_\G (\theta - \H dt) =  \theta - \H dt = \L \ee
and hence although the action is rescaled by an overall constant, the conditions that its minimization places upon the invariants is unchanged. In other words, the equations of motion of the invariants of $\G$ are unaffected by the action of $\G$. The boundary term introduced does not consist of invariants of $\G$. 

A direct corollary of this is the autonomy of invariants of $\G$: Given a set of invariants of $\G$, the relative evolution of any two invariants is itself invariant. Evolution relative to an invariant is given by the scaled evolution along the Hamiltonian vector field $\X_\H$. 
\be \Lie_\G \left(\frac{d\State_1}{d\State_2} \right) =  \Lie_\G \left(\frac{\dot{\State_1}}{\dot{\State_2}}\right) =  \Lie_\G \left(\frac{\iota_{d\State_1} \X_\H}{\iota_{d\State_2}\X_\H } \right) = 0 \ee
and thus the relational motion must also be an invariant. The system's closure is inherited from the closure of the dynamics of the Lagrangian. The orbit of the a generator of dynamical similarity forms a one dimensional subspace of the phase space under which invariant dynamics is unaffected.  Since the generator of dynamical similarity acts to rescale the Lagrangian by an overall factor, the equations of motion of the invariants are unaffected, and these invariants form a closed system. An observer who had access to all the invariants of the theory could derive the entire evolution of the system without ever referencing the non-invariants.

We note from this that not only does the existence of a dynamical similarity necessarily imply the existence of an invariant system, but also informs us as to the form of the invariant variables themselves. These are functions of the coordinates and momenta in phase space, $f(q_i, p^i)$ such that $\Lie_\G f = 0$. An arbitrary power of an invariant is also invariant, as any linear combination of invariants. Given any invariant coordinates (or momenta), the ratio of their conjugate momenta (coordinates) will also be invariant. Further, given any two non-invariants basis elements of phase space, $x,y$ with eigenvalues $a,b$ respectively, the ratio $\frac{x^b}{y^a}$ is also invariant. From this and Leibniz rule, it is apparent that invariants form an algebra. 

Let us here recapitulate the significant results of this section. We have shown a general form for a non-strictly canonical transformation which acts on phase space variables such the the Hamiltonian is rescaled. Using this form we established that there exists an algebra of invariants, whose evolution is unaffected by the transformation and autonomous. Therefore given such a dynamical similarity, an observer who only had access to relational degrees of freedom could not distinguish where along the orbit of the symmetry they were. An intrinsic observer would identify the same physics in each situation.
 
\section{Contact Forms and Shape Space} \label{Contact}
 
\noindent Once we have determined the existence of a dynamical similarity and identified the algebra of invariants, our goal is now to describe physics purely in terms of these invariants. Since these form an autonomous system within the full symplectic framework, we could express dynamics in the full framework and project down to the invariants, using the unobservable directions like Wittgenstein's ladder. However, it is possible that the extended dynamical system will have points at which the equations become divergent due to the behaviour of the unobservables. This was shown to be the case in Homogeneous cosmology recently, where it was found that although singularities exist in the extended framework, there exists an autonomous subset of invariant variables whose evolution remains well defined even at the big bang singularity \cite{Koslowski:2016hds}. Therefore we are motivated to construct our physical theory directly on the space of invariants and show that this gives equivalent evolution for the observables to the symplectic system where the latter remains well defined. This motivates the introduction of `contact dynamics' - a counterpart to symplectic dynamics that takes an odd dimensional space as the basis for physics \cite{ArnoldBook}. Our goal in this section is to show the equivalence between a symplectic system with a dynamical similarity and contact dynamics of the reduced space of invariants.  

We begin with an odd-dimensional manifold, $\M$ with dimension $n+1$. A contact form $\eta$ on this space is the odd-dimesional counterpart to the symplectic potential \cite{ContactIntro}. In particular, $\eta \wedge (d\eta)^{n/2}$ is a volume form on $\M$. There exists universally a set of coordinates $A, y^i, x_i$ for $\M$, the `Darboux coordinates', in which $\eta = -dA + y^i dx_i$. Primarily, contact geometry is concerned with the kernel of $\eta$, and hence it is normally defined up to an overall scale \cite{geiges2008introduction}. This freedom to rescale $\eta$ is important when recovering the symplectic system from which the contact dynamics is derived.

Given a vector field $\X$ on $\M$, the contact Hamiltonian, $\H^c$ is $\iota_\X \eta$ \cite{Contact1}. The relationship with the usual Hamiltonian on an even dimensional space is that the contact Hamiltonian together with the contact form generate the flow. In Darboux coordinates the dynamics of the system is then given:
\be \dot{x}^i = \frac{\partial \H^c}{\partial y_i} \quad \dot{y_i} = -\frac{\partial \H^c}{\partial x_i} - y^i \frac{\partial \H^c}{\partial A} \quad \dot{A}= y_i \frac{\partial \H^c}{\partial y_i} - \H^c \ee
It can then be easily shown that $\dot{\H^c} = -\H^c  \frac{\partial \H^c}{\partial A}$. Here we see an important distinction from the symplectic dynamics - the contact Hamiltonian is only conserved when it is either independent of $A$ (and hence $\frac{\partial}{\partial A}$ becomes a symmetry of the system) or zero explicitly. Hence the dynamics of a contact system is similar to that of a nonconservative flow, and this system will exhibit friction-like properties.  

The relation with our interest in dynamical similarity is readily apparent. Treated as a manifold, the set of invariants of a generator of dynamical similarity $\G$ form a contact manifold with contact form $\eta= \frac{\iota_\G \omega}{\rho}$ in which $\rho$ is an eigenfunction of $\G$ with eigenvalue 1. Chosen in this way, $\eta$ is an invariant of $\G$:
\be \Lie_\G \eta = \frac{\Lie_\G \iota_\G \omega}{\rho} - \frac{\Lie_\G \rho}{\rho^2} \iota_\G \omega = \frac{\iota_\G d\eta}{\rho} - \eta =0 \ee
wherein we used the facts that $\iota_\D^2=0$ and $d\eta=\omega$. Going in the other direction is also quite simple; the space of contact forms constitutes a one-dimensional fibre bundle over $\M$, with contact forms related to one another by multiplication by a positive real number. Thus we can form the symplectification of our contact system by first expressing the contact form in (local) Darboux coordinates, and promoting the choice of scalar to a coordinate. Note that since the choice of scalar for the contact form is global, we can unambiguously promote it to a global coordinate over the contact manifold. Thus expressed, $\rho \eta = -\rho dA + \rho y^i dx_i$. We further identify coordinates on this even dimensional manifold: $p^o = A , p^i = \rho y^i, q_o =\rho, q_i = x_i$ to obtain a symplectic potential of the usual form up to an exact form: 
\be \theta = -q_o dp^o + p^i dq^i = p_o dq^o + p^i dq^i - d(p^o q_o) \ee
which corresponds to $\iota_\D \omega$ where $\omega=d\theta$ is the usual symplectic structure, and 
\be \D = q_o \frac{\partial}{\partial q_o} + p^i \frac{\partial}{\partial p^i} \ee
The contact Hamiltonian is the restriction of the Hamiltonian to the invariants in a particular choice of lapse:
\be \H^c = \iota_\eta \X_\H = \X_\H(\rho \iota_\D \omega) = \rho^{-1} \omega^{-1} (\iota_\D \omega, d\H) = \rho^{-1} \iota_\D d\H = \rho^{-1} \Lie_\D H = \rho^{-1} \Lambda \H \ee
and hence on such systems the dynamics generated by the contact Hamiltonian will agree exactly with the dynamics of the invariants of $\D$. Furthermore, a specific choice of lapse, allows us to pick a time coordinate such that $\H^c \propto \frac{\H}{\rho^\Lambda}$ in which case we see an important result: the contact Hamiltonian belongs to the algebra of invariants of $\D$. The operations of differentiation is closed on the algebra, hence we see that the derived contact dynamics of this system will indeed be simply a function of the invariants alone. Hence we see that a zero energy Lagrangian system with a dynamical similarity is equivalent to a contact system on the invariants of the dynamical similarity, and vice-versa. The symplectification of a contact system results in a dynamical system with the same equations of motion on invariants and has a dynamical similarity which can be used to restore the contact system. We can therefore translate freely between contact systems and symplectic systems in this framework. An identical result holds when the Lagrangian is homogeneous in one of the variables - see below. 

In the majority of cases the condition that $\H$ is constant will allow us to solve for $A$ as a function of the $y^i$ and $x_i$, and the energy of the system. More precisely, if the surface $\H^c=0$ is covered in some local chart by $A=g(x,y)$ for some (possibly multivalued) function $g$, then this requirement is just that $\frac{\partial \H^c}{\partial A}$ restricted to the $\H^c=0$ surface is independent of the choice of branch of $g$. This is not greatly restrictive and often will amount to a choice of sign for a square root, for example. 
In such a system, then, the entire dynamics will be expressed only in terms of the `shapes' $x_i$ and their velocities $y^i$, with some initial choice of $E$. Since time is only determined up a lapse in the system, this means that the $\dot{x_i}$ are only required up to scale; a point and a direction on the shapes is enough to determine an evolution.

The contact form determines a volume form $Vol=\eta \wedge d\eta^{n/2}$ on the system. In the case where $\frac{\partial^2 \H^c}{\partial A \partial y}=0$ the evolution of this volume form has a very simple form. Note that this condition is not particularly restrictive; in terms of the original symplectic system it amounts to the kinetic term being diagonalizable in momenta. Since $\iota_\eta \X_{\H} = \H^c$ we see that 
\be \Lie_{\X_\H} \eta = -\frac{\partial \H^c}{\partial A} \eta \ee
Since the Lie and exterior derivatives commute, we can thus calculated the action on the volume form, $Vol$ The Lie derivative of this along the Hamiltonian flow is then \cite{ContactLiouville}
\be \Lie_\X Vol = - (n/2+1) \frac{\partial \H^c}{\partial A} Vol \label{Divergence} \ee
which determines the divergence of any flow. We see that this divergence is exactly zero when $\frac{\partial \H^c}{\partial A}=0$ - this is a Janus point, a point along a trajectory where the volume form would turn around. At this point, the dynamics of the system is instantaneously equivalent to that of a symplectic system in which $x$ and $y$ are a conjugate pair evolving under the contact Hamiltonian with $A$ treated as a constant. Following the arguments employed by Barbour et al. \cite{Barbour:2014bga,Barbour:2013jya}, this is a necessary prerequisite for an ``arrow of time" to emerge in such systems - it's the direction of focussing of the contact form. For an intrinsic system, the notion of time is complex; it requires the formation of 'records' - information about past configurations which is available to a contemporary observer. As such these are expected to exist on a lower dimensional subspace of the space of shapes. This is a subtle issue, with wide-ranging philosophical implications which is discussed at length in \cite{Gomes:2010fh,Gryb:2014mva,Gomes:2016aam,Evans:2016org}. Furthermore care should be taken to distinguish between a variable which is to be used as a clock, and one from which chronology can be established. In general, a chronology is established to provide an ordering on a set of events. A clock further can be used to define a unit of time. As noted in the introduction, the relevance of using, he rotation of the Earth as a clock is that the periodicity of the system is (approximately) proportional to the periodicity of the orbit of the Earth around the sun. The use of a massless scalar field in cosmological models, however, provides a chronology but would be of little further use as a clock. In establishing a chronology for our systems, of particular import is that the contact version of Liouville's theorem is details the behaviour of the volume form alone, any measure expressed in invariants can be used together with this volume form without qualitatively changing the focussing result. Note that this apparent direction of time may differ from the configuration time $t$ which we have used as a parameter along solution curves. When applied to the universe as a complete system, this has significant implications for the `Past Hypothesis', which is highly contentious \cite{EARMAN2006399}. Since the odd dimensional form of Liouville's theorem does not conserve volume forms, the fact that there is a state of low entropy in our past is no longer surprising, but in fact a natural consequence of the dynamics. The usual arguments regarding the Past Hypothesis are all based in physics with conserved volume forms, however as our dynamics will break time symmetry as experienced by an intrinsic observer, such arguments need to be reconsidered.

\section{Homogeneous Lagrangians} \label{Homogeneous}

\noindent It is now easy to describe a simple yet interesting class of Lagrangians - those which are homogeneous in one of the configuration variables. Under a simple canonical transformation $Q= q^n$ we see that the degree of homogeneity is immaterial as such transformations can be used to always pick a variable in which the degree is 1, for example. Such Lagrangians encompass minimally coupled gravitational systems as example cases (see the attractor papers for the early work on this). Therefore suppose that a Lagrangian, $\L_h$, is homogeneous of degree 1 in the configuration variable $x$, i.e. for some $\alpha \in \mathcal{R}$
\be \L (\alpha x, \alpha \dot{x}, \vec{q},\dot{\vec{q}}) = \alpha \L (x, \dot{x}, \vec{q},\dot{\vec{q}}) . \ee
The Euler-Lagrange equations for the $q_i$ are unaffected by this transformation. Within such a system, $x$ cannot be a member of any shape space, as the dynamics is insensitive to the value of $x$ up to this overall choice of scale. A trivial calculation shows that $P^x$, the momentum conjugate to $x$ is unaffected by this change, nor is the equation of motion for $P^x$. However, the momenta conjugate to the $q_i$ are rescaled by $\alpha^{-1}$. Therefore we see that at the phase space level, to reproduce this transformation we must rescale the symplectic potential by the same factor, $\alpha$. Thus the dynamical similarity is 
\be \F = x \frac{d}{dx} + P^i \frac{d}{dP^i} .\ee
A simple calculation quickly reveals that this is indeed a dynamical similarity:
\be \iota_\F \omega = P^i dq_i - x dP^x = \theta - d \circ \iota_\F \theta \quad \Lie_\F \H_h = \H_h \ee
A much more laborious direct calculation reveals that $\Lie_\F \X_\H = 0$. Further, under the action of $\F$, $P^x$ is an invariant, as are the $xP^i$. 

In essence this is the reason that the Friedmann equation in homogeneous, flat cosmology is independent of the volume $v$, but does depend on its conjugate momentum, which is the Hubble parameter. 

In these systems, the symplectic form on phase space naturally induces a contact form on the space of invariants, $\eta=\frac{\iota_\F \omega}{x}$. Since the Hamiltonian flow is independent of the position along the orbit of $\F$, we find the contact Hamiltonian is the usual Hamiltonian
\be \H^c = \X_\H (\eta) = v^{-1} \left(x \frac{d\H}{dx} + P^i \frac{d\H}{dP^i} \right)= v^{-1} \Lie_\F \H = \frac{\H}{v}\ee
The contact form can be expressed in terms of the space of invariants of $\F: \{P^x, q_i, p^i=\frac{P^i}{x}\}$. In terms of these invariants, $\eta = x(p^i dq_i - dP^x)$ and $\H=x h(P^x, q_i, p^i)$. Thus any Lagrangian which is homogeneous in one of the configuration variables, $X$ can be naturally associated with a contact system on the reduced phase space consisting of the conjugate momentum to $X$, the remaining configuration variables and their conjugate momenta divided by $X$. 

An important subset of these Lagrangians is are those describing the dynamics of flat Robertson-Walker cosmologies minimally coupled to matter. Such cosmological models consist of a mini-superspace model in which the only gravitational degree of freedom corresponds to the volume $v$ of a fiducial cell. This is usually expressed through the scale factor $a=v^{1/3}$. Here we will remain in the volume representation for two reasons. The first is that in General Relativity the Hubble parameter is the conjugate momentum to volume. The second is that in these variables it is readily apparent that minimal coupling of gravity to matter comprises a homogeneous Lagrangian. The line element is then
\be ds^2 = -dt^2 + v(t)^{2/3} \left(dx^2 + dy^2 + dz^2 \right) \ee
and our phase space consists of the geometrical variables $v, P^v$ and the matter degrees of freedom $q_i p^i$. The nature of the gravitational action will relate these and provide a Hamiltonian from which dynamics can be determined. Any action based solely on geometrical quantities (or equivalently, one which does not introduce an external notion of scale) such as the Ricci tensor must be homogeneous in $v$ since the theory is independent of the choice of fiducial cell used to determine $v$. 

These were discussed extensively in \cite{Corichi:2013kua,Sloan:2014jra} with the existence of attractors established within the broader framework of the symplectic structure \cite{Sloan:2016nnx}. The general structure of a gravitational action minimally coupled to a matter Lagrangian $\L_m$ with symplectic structure $\omega_m$ is in such cases
\be \L = v\left(f(\frac{\dot{v}}{v}) + \L_m [q,\dot{q}] \right) \ee
and hence the homogeneity of the Lagrangian in $v$ is readily apparent. From the dynamical similarity of this system we can form the contact form $\eta = -dP^v + P^i dq_i$ wherein $P^i$ are the momenta of the matter Lagrangian treated as free from interaction. Therefore the contact form  is the symplectic potential for the uncoupled matter system, added to the exact form along the remaining orthogonal direction,  $\eta = \theta_m - dP^v$. The contact Hamiltonian is the Hamiltonian of the matter component added to a function of the Hubble parameter which acts as friction:
\be \H^c = F(P^v) + \H_m  \ee
Note that for example in General Relativity, $F(P^v) \propto {P^v}^2$ and $H=P^v$, in Loop Quantum Cosmology $F(P^v) \propto \sin^2(\frac{P^v}{\Delta})$ \cite{Ashtekar:2012np}etc. We can now find the dynamics of our system in terms of the positions and momenta of the matter system. Since the contact form is already in Darboux coordinates we find:
\be \dot{q_i} = \frac{\partial \H_m}{\partial P^i} \quad \dot{P^i} =\frac{\partial \H_m}{\partial q_i} - P^i \frac{\partial F}{\partial P^v} \quad  \dot{P^v} = \sum P^i \frac{\partial \H_m}{\partial P^i} \ee
From which we see a clear physical parallel for the attractors of this system. The coupling to gravity makes the matter system behave as though it were subject to friction terms - it is a dissipative system on the shape space. The presence of attractors is therefore unsurprising; the expansion of the universe removes energy from the matter system. Since $F$ is a function which is independent of the matter degrees of freedom, so is its derivative with respect to $H$. Up to a possible choice of branch, $F'$ can be inverted, and solving the Hamiltonian constraint will allow us to express the frictional term encountered as a function of the matter degrees of freedom alone. 

The volume form is $dH\wedge Vol_m$ in which $Vol_m = \omega_m^n$ is the volume form on the matter phase space. From equation (\ref{Divergence}) we see that the focussing of the volume form comes from the expansion of the universe, giving a natural volume weighting \cite{Sloan:2015bha}:
\be Vol(t_2) = Vol(t_1) \exp [\int Hdt] = \frac{v_2}{v_1} Vol (t_1) \ee
Thus solutions with the greatest expansion are attractors. It further follows that any point of bounce or recollapse of a solution is a Janus point, regardless of the specific gravitational theory, and that the ``arrow of time" in the Barbour sense, must point in the direction of the expansion of space. The space of solutions to the contact Hamiltonian constraint is finite for any given choice of the Hubble parameter, and thus evades the serious measure problems inherent in non-compact spaces \cite{Curiel:2015oea}. This is a direct result of the dynamical similarity, which has reduced the space of solution by identifying those in the symplectic system which are connected by the orbits of the dynamical similarity. In this case, the symplectic system is non-compact, as the choice of volume at any given Hubble parameter is restricted to the positive real line. However, the contact system is insensitive to these changes, and thus the space of intrinsically distinguishable solutions is compact, and hence we avoid significant topological issues. This was first identified in \cite{Measure,Corichi:2010zp,Measure2} in the context of inflationary models in Loop Quantum Cosmology, wherein the dynamics provided a natural bounce point at which to evaluate this measure.

\section{Dilations in the n-body system} \label{NBody}

\noindent We will here examine the case of dynamical similarity within a systems defined by a single Lagrangian with fixed external parameters. For clarity, as we will be using powers of momenta and coordinates regularly, we will write both coordinates and momenta with their indices lowered henceforth. As we have shown, within such a system if the phase space is of higher dimension than the space of physical observables there will be redundancies in the description. Those redundancies which generate strictly canonical transformations are the usual gauge symmetries, and those which generate their non-strict counterparts are the dynamical similarities. One such example of a system which has such a dynamical similarity is the shape space of an $N$ body system in $d$ dimensions. Intrinsically, we have no access to a rod with which to measure the separation of any two particles in this system, therefore we will identify any two configurations that are related by a rescaling of all the distances between particles. Thus the generator of dilations will act to provide the dynamical similarity. In particular, consider a systems described in cartesian coordinates by 
\be \L =\frac{\dot{\vec{q_i}}^2}{2} - V(\vec{q_1},...,\vec{q_N}) \ee
wherein to potential $V$ is homogeneous of degree $\gamma$: $V(\lambda \vec{q}) = \lambda^\gamma V(\vec{q})$, and Gallilean invariant: $V(\vec{q_1}+\vec{x},...,\vec{q_N}+\vec{x}) = V(\vec{q_1},...,\vec{q_N})$ wherein $\vec{x}$ is a uniform translation of all the particle positions (i.e. a change of origin for our system) and $V(M\vec{q_1},...,M\vec{q_n}) = V(\vec{q_1},...,\vec{q_N})$ wherein $M \in O(d)$ acts on the position vectors of the particles identically. Potentials, such as the Coulomb potential, which only depend on particle separations $V=V(\sum_{i \neq j} C_{ij} |\vec{q}_i - \vec{q}_j|^n)$ are members of this type. It is immediately apparent that the generators of these transformations, $\mathcal{T}$ and $\mathcal{O}$ are symplectomorphisms, and their associated N\"other charges (momentum and angular momentum) are often used to reduce the description to center of mass coordinates with zero net angular momentum. \footnote{Technically it is the continuous transformations of one of the two subgroups $SO(n)$ that generates angular momentum, and the N\"other charge relies only on the continuous subgroup. Reflection is a discrete change and therefore has no associtate N\"other current. Other discrete symmetries such as the interchange of two particles will be important in defining statistical ensembles (Fermi vs Bose statistics).}

To form the shape space of this system we need to identify a (function of a) (subset of the) coordinate(s) to use as a rod. Ideally this will be done such that the Hamiltonian flow on phase space is easy to pull back onto the shape space. For simplicity of exposition we will assume a frame in which the system has zero net angular momentum, but the center of mass of the system has position $\vec{o}$. At this point we retain this degree of freedom simply to show the difference between gauge identifications under the generator of translations of the system (strictly canonical) and that of dilations (non-strictly canonical). 

The symmetry due to translation of the center of mass is $\Delta_o = \frac{d}{do}$, and we can see clearly that this generates a canonical transformation as $\iota_{\Delta_o} \omega = dP_o$ which is clearly exact. Since the Hamiltonian is independent of the choice of origin of coordinates, $\Lie_{\Delta_o} \H=0$. The corresponding freedom of choice frame under changing the center of mass momentum is generated by $\Delta_{P_o} = \frac{\partial}{\partial P_o}$ which is also a symplectomorphism, and acts to shift the Hamiltonian by a constant. This should be unsurprising as we are removing energy from the system in transforming to this coordinate basis. Since the change to the Hamiltonian is a constant, the pullback of the Hamiltonian flow onto shape space is unaltered. We can therefore unambiguously project our system onto a subspace of the original phase space defined by $o=P_o=0$, making the obvious pullbacks of the symplectic structure and Hamiltonian (noting that the first term in $\theta$ is trivially zero). 

We also need to define a rod. A democratic choice of such is to define a length scale by $R^2 = \vec{q}^2$. Our choice of evolution of scale will be such that $R$ is fixed in time, and typically we will choose $R=1$. Thus we find that the shape space has dimension $D_s=2d(N-2)-2$ and is the sphere $S^{D_s-1}$. The scaling of the potential is explicit here: $V=R^n V_s$ wherein $V_s$, termed the shape potential only depends on coordinates on the shape sphere. If $n=2$ we would have a homogeneous Lagrangian, and this would fall into the category discussed above. If $n=-2$ the system is conformally invariant, and thus dilations are strictly canonical. The symmetry under dilation is already well understood, and thus we exclude this from further discussion. For convenience of making explicit the act of the symmetry under choice of $R$ we can express the Lagrangian in terms of this system as
\be \L =  \frac{\dot{R}^2}{2} + \frac{R^2 \dot{\vec{T}}^2}{2} - R^n V_s [\vec{T}] \ee
wherein we have introduced $T[\vec{\phi},\vec{\dot{\phi}}]$ to represent the trigonometric functions which describe the positions on the shape sphere corresponding to the particle positions. These are the functions 
\ba T_i &=& \sin(\phi_1)\sin(\phi_2)... \cos(\phi_i) \quad i<D_s \nonumber \\
      T_{D_s} &=& \sin(\phi_1)\sin(\phi_2)...\sin(\phi_{D_s})\ea
Naturally $\sum T_i^2=1$, and it is trivial, but tedious, to show the $T_i$ and their derivatives are are mutually orthogonal and all are orthogonal to $R$. Henceforth we will drop the vector notation from $\vec{T}$ to avoid symbolic clutter. From our setup we are now ready to perform the usual Legendre transformation and obtain a Hamiltonian form and symplectic structure:
\ba \H 	&=& \frac{P_R^2}{2} + \frac{P_T^2}{2R^2} + R^n V_s(T) \nonumber \\ 
\omega	&=& dP_R \wedge dR + dP_T \wedge dT \ea
wherein we have obtained $\omega$ from the symplectic potential $\theta = P_R dR + P_T dT$. 
Dilations of our system should map rescale $R$, yet leave the dynamics unchanged. The vector field generating dilations is (up to a choice of scale) thus
\be \D =\frac{1}{n+2} \left( 2 R \frac{\partial}{\partial R} + (n+2) P_T \frac{\partial}{\partial P_T} + n P_R \frac{\partial}{\partial P_R} \right) .\ee
which we see has the correct action on interior product with the symplectic two-form:
\be \iota_\D \omega = \theta - \frac{2}{n+2} d(RP_R) =\theta - d\circ \iota_\D \theta \ee
and $\Lie_\D \H = \frac{2n}{n+2} \H$. 

The boundary states of our system, $\State_1$ and $\State_2$, will depend only on the shape variables, $T_i$. Thus our time parameter $t$ must be constructed from the orbits of these. We note that $P_T = R^2 \dot{T}$ and thus we know that
\be dt = \frac{R^2}{P_T} dT \ee
for any $T$ chosen from the $T_i$. Since these are shape variables, $\D$ does not act upon them, and thus we calculate
\be \Lie_\D dt = \iota_\D \circ d (\frac{R^2}{P_T} dT) = \frac{2-n}{n+2} dt \ee
and hence we see that the Hamiltonian as a one-form is Lie dragged exactly, using Leibniz rule:
\be \Lie_\D \H dt = \frac{2n}{n+2} \H dt + \frac{2-n}{n+2} \H dt = \H dt \ee
We will now construct the invariants from which the autonomous dynamics of the system can be expressed directly. Since we have not specified $V$, and have explicitly chosen $\D$ to be orthogonal to the coordinates $T$, these will constitute one of the invariants. We note that $\rho = \frac{2}{n+2} R^{\frac{n+2}{2}}$ is an eigenfunction of $\D$ with eigenvalue 1, hence can be used to to find the invariants. By focussing on the action of $\D$ on $R$ we then see that a construction of the two remaining invariants from $R$ and the momenta $P_R$ and $P_T$ gives us our compete set:
\be A=\frac{P_R}{R^{n/2}} \quad B=\frac{(n+2)}{2} \frac{P_T}{R^{\frac{n+2}{2}}} \ee
wherein we have chosen the prefactor of $B$ such that the invariants are Darboux coordinates for the contact form $\eta$:
\be \eta =  \frac{\iota_\D \omega}{\rho} =  BdT - dA \ee
In this formulation, the contact Hamiltonian is then, up to a choice of lapse:
\be \H^c = \frac{A^2}{2} + \left(\frac{2}{n+2}\right)^2 \sum_i \frac{B_i^2}{2} + V_s \ee
from which we obtain the contact equations of motion: 
\be \label{ShapeEom} A' = \left(\frac{2}{n+2}\right)^2 \sum_i  B_i^2 \quad 
      B_i' = -\frac{\partial V_s}{\partial T_i} - AB_i \quad 
      T_i' = \left(\frac{2}{n+2}\right)^2 B_i \ee
Let us here point out explicitly that $A$ is monotonically increasing, and outside a set of measure zero under the measure induced by the contact form in which $B_i=0$ for all $i$ on constant potentials, there will be a unique point at which $A=0$ on each trajectory. Per our earlier definition, this is the \textit{Janus point} of the system, the point where $\frac{\partial \H^c}{\partial A} = 0$.  From the conservation of the contact Hamiltonian we can eliminate $A$ from the system, and render the complete dynamics as a set of second order ordinary differential equations, one for each angular position $T_i$:
\be \ddot{T_i} +\dot{T_i} \sqrt{-(2V_s +\left(\frac{n+2}{2} \right)^2 \sum \dot{T_j}^2)}  + \left(\frac{2}{n+2} \right)^2 \frac{\partial V_s}{\partial_{T_i}} =0 \ee
which makes clear that dynamics is determined by a point and a direction in shape space. 

Let us now establish the relationship between the description here, and the shape space description given by Barbour et al. In their seminal work, they established the shape dynamics of an N-body system subject to a Newtonian potential. In our terminology, this is the case in which $n=-1$ and $V_s = -\sum_{a<b} (T_a-T_b)^{-1}$. In this case, the Janus point was identified at the point at which the dilatational momentum vanishes. This dilatational momentum corresponds to $A$, and we see that the Janus point that we have identified is exactly that described; $\frac{\partial\H^c}{\partial A} = 0 \rightarrow A=0$. 

To understand the eventual behaviour of our system first note that since $A$ is monotonically non-decreasing on solutions, the form of the contact Hamiltonian means that a solution must be either slowing ($\sum \ddot{T_i} <0$) or heading down the potential. There exists a stationary configuration: $\dot{B_i}=0= \dot{T_i}$ wherein the particles are all equidistant from one another on shape space, and thus $\frac{\partial V_s}{\partial T_i} = 0$ The systems split into two groups depending on the sign of $n$; if this is positive the stationary solution is stable. However if $n$ is negative then the stationary solution is unstable to small perturbations. In this case the solution will always seek out the (infinitely deep) wells of the potential which correspond to local, isolated, trapped systems.

\section{Homogeneous Cosmology} \label{Cosmology}

\noindent In recent work \cite{Koslowski:2016hds} the shape dynamics of a Bianchi IX system was examined. It was shown that there is an autonomous subsystem of dynamics that arises in terms of the shape variables, and that this system remains deterministic through the singularity. Here we will show that this is achieved in part due to the dynamical similarity that is present in such systems; as we have already established the existence of such a similarity implies the existence of the autonomous subsystem. One key difference from the dynamics of General Relativity is that the volume of the universe (and its conjugate momentum, the Hubble parameter) is not a member of the algebra of invariants. Therefore in constructing a geometrical representation of the theory, further external inputs are required which play no role in the evolution of the invariants. Thus, although the geometrical picture breaks down at a singularity, it was found that the dynamics of these invariants does not, and beyond the point at which GR is singular a geometrical picture can be redeveloped from the invariants. A similar phenomenon in the evolution of geodesics through a Schwarzschild black hole has been recently discussed \cite{Bianchi:2018mml,CarloPrep} which is indicative that this may be a more general property of Einstein's equations. 

In this section we will show how dynamical similarity leads to an autonomous system in cosmology. The metric for a homogeneous (but possibly anisotropic) space-time can be expressed in terms of the translation invariant one-forms $\sigma_i$ on the spatial manifold, $\Sigma$
\be ds^2 = -dt^2 + v^{\frac{1}{3}} \exp(\gamma_i) d\sigma_i \ee
wherein $v$ labels a choice of volume of the $\Sigma$, and the $\gamma_i$ are anisotropy parameters. These are constrained such that their sum is zero, and thus can be spanned by the two Misner parameters:
\be \gamma_1 = -q_1/\sqrt{6} - q_2/\sqrt{2} \: \gamma_2 = -q_1/\sqrt{6} + q_2/\sqrt{2} \: \gamma_3 = \sqrt{\frac{2}{3}} q_2 \ee
Matter in the form of a massless scalar field ($\phi,\pi$) is simple to add to the system through minimal coupling. The symplectic structure is then:
\be \omega = dp_i \wedge dq_i + d\pi \wedge d\phi + d\tau \wedge dv \ee
 and the ADM Hamiltonian is given:
\be \H = -\frac{3}{8} v^2 \tau^2 +p_1^2 + p_2^2 + \frac{\pi^2}{2} +  v^{\frac{4}{3}} V_s (q_1,q_2) \ee
Wherein $V_s$ is the shape potential, and $v^{4/3} V_s$ is the Ricci scalar of $\Sigma$, which in turn is determined by the algebra of commutativity of the $\sigma_i$. In the case of a flat space-time (Bianchi I), $V_s=0$, and the dynamics is quite simple. However in general the 3-geometry can be more complex, with topologies of e.g. $S^3$ (Bianchi IX) or $S^1 \times S^2$ (Kantowski-Sachs), in which case $V_s$ has a more complicated form. Therefore we will restrict our analysis to vector fields that leave $q_1$ and $q_2$ (and hence $V_s$) invariant. A simple direct application shows that 
\be \G = p_i \frac{\partial}{\partial p_i} + \pi \frac{\partial}{\partial \pi} - \frac{\tau}{2} \frac{\partial}{\partial \tau} + \frac{3v}{2} \frac{\partial}{\partial v} \ee
is a dynamical similarity. Hence there is an autonomous subsystem of invariants of $\G$. Since we explicitly chose $\G$ to preserve the Misner coordinates, $q_1$ and $q_2$, these are two of the invariants of $\G$. Similarly, since $\H$ is independent of $\phi$ (which is an invariant), $\pi$ is a constant, and the value of $\phi$ does not affect dynamics. From the scaling of $v$, we can form a set of invariants: 
\be \psi_i = \frac{p_i}{v^{2/3}} \quad \psi_\phi = \frac{\pi}{v^{2/3}}  \quad \Phi = \frac{3}{2}\tau v^{1/3} \ee
Note that these variables differ slightly from those used in \cite{Koslowski:2016hds} - we have chosen these such that the contact system is simple to write in Darboux coordinates. Thus we have a 7 dimensional space of invariants whose dynamics close. Further, the value of $\phi$ does not contribute to the dynamics of the other invariants, as the Hamiltonian is independent of $\phi$, and the Hamiltonian constraint can be used to eliminate a further invariant, hence a set of 5 invariants, independent of $\phi$, form a closed dynamical system. The contact form is then:
\be \eta = \frac{\iota_G \omega}{v^{2/3}} = \Psi_i dq_i + \Psi_\phi d\phi- d\Phi \ee
and the contact Hamiltonian is
\be \H^c = \Psi_1^2 + \Psi_2^2 + \frac{\Psi_\phi^2}{2} - \frac{\Phi^2}{6} + V_s (q_1, q_2) \ee
Note that our contact Hamiltonian contains six variables, however it is a constrained to be exactly zero, and hence there is only a five dimensional space of solutions. The equations of motion for our system in terms of the invariants are then:
\ba \dot{q_i}&=&2\Psi_i \quad \quad \dot{\phi} = \Psi_\phi \quad \dot{\Psi_i} = -\frac{\partial V_s}{\partial q_i} - 2 \Phi \Psi_i \nonumber \\
 \dot{\Psi_\phi}&=&-\Phi \Psi_\phi \quad \dot{\Phi}=2\Psi_1^2 + 2\Psi_2^2 + \Psi_\phi^2 \ea
Thus we have constructed the autonomous system of invariants which describes homogeneous cosmology. In \cite{Koslowski:2016hds} this system is shown to remain predictive beyond the singularity. The analysis there requires some coordinate transformations, and a direct investigation of the regularity of the differential equations to show that they are predictive at the singularity. 

\section{Discussion} \label{Discussion}

\noindent In this work, we have established three fundamental facts about dynamical similarity. These are: the existence of a vector field on phase space which generate the symmetry, the closure of the system of invariants and the relationship with contact dynamics on the space of invariants.

We have given a prescription for finding the invariant dynamics of a symplectic system with a dynamical similarity. Such symmetries are revealed by the existence of a non-strictly canonical transformation that is also Hamiltonian scaling. The existence of such symmetries should not in fact be surprising; in many cases they correspond to simply an arbitrary choice of, for example, a unit of length within a system. It is therefore to be expected that altering this choice of unit should not affect the intrinsic, relational dynamics. Thus there is a redundancy in the phase space description corresponding to this choice. This is similar in many ways to other symmetries of such systems. The freedom to choose a reference frame in particle systems also does not affect intrinsic dynamics. Likewise, the remaining phase space variables form an autonomous system; their relational evolution can be expressed without referring to the choice of frame. However within the Hamiltonian formalism, such choices correspond to symplectomorphisms and are strictly canonical, which in turn contribute boundary terms to the action. The key difference for dynamical similarities is that these also rescale the symplectic structure. Thus we see that such transformations may alter the conserved quantities of a system. However, this indicates that to an observer who only has access to the intrinsic observables of the system such the changes resulting from such transformations are not measurable.

A complementary approach to that discussed here is developed in \cite{TimPrep}, in which the theory is directly constructed taking the intrinsic observables as fundamental. This begins from two fundamental postulates. The first is that the phase space of the system consists of the smallest possible set of geometric parameters required to close an equation describing a curve through shape space. The second is that the equation of state of the curve arises from the (unit) tangent bundle over this phase space. In such a construction one begins with intrinsic observations and forms equations of motion directly, arriving at the contact systems which were discussed in section \ref{Contact}. This is done explicitly in the case of the three body problem, which is the simplest system which has non-trivial intrinsic dynamics. To see this directly, consider that the two body problem can be expressed in center-of-mass coordinates, and reduces to a single body in an external potential. Once the separation of the two bodies is used to define a rod, there can be no further dynamics of the system.

In the case of Newtonian gravity, the construction is made explicit, and it is shown how the experienced space-time can be reconstructed by an observer who makes certain necessary choices of scales in order to embed the relational system within a system with absolute notions of scale. The emergence of isolated systems (particularly Kepler pairs) which can be used as de-facto rods and clocks is shown explicitly. This construction is entirely compatible with that expressed in this paper, and the embedding within a symplectic system in essence is the promotion of the scale to a dynamical variable. Thus by construction the resultant system will have a dynamical similarity which corresponds to the choice of such scale, and the processes outlined here will necessarily recover the original relational system. 

We have shown that dynamical similarity reveals the underlying structure of a symplectic system, which is a contact system which yields dynamics in terms of the invariants of the transformation alone. The expression of dynamics in these terms was a key part of the continuation past singularities in \cite{Koslowski:2016hds}, and the principle factor behind showing that there was a physically well determined volume form on which measures could be based in \cite{Sloan:2015bha}. The existence of an intrinsically defined arrow of time necessarily requires some degree of focussing of dynamics such that records can be formed. The friction-like terms which arise in this formalism are central to this realisation, as under the dynamical flow the volume form is not conserved. Thus dynamical similarity within a symplectic system provides some of the ingredients from which a full intrinsic theory can be constructed.

In the present construction we have taken into account only the dynamical similarities which are generated by acting on phase space variables. This was done so that the basic mathematical structure was direct to establish, both in terms of the vector fields that generate them, and the contact structure which was found to underlie the intrinsic dynamics of invariants. However, a more general formulation will also include the effect of altering coupling coefficients in the systems themselves. In formulating a Lagrangian, the strengths of couplings must be given explicitly in the construction of the theory. However, in practice these values are established by fitting observations of relational variables. Therefore the values of the couplings themselves must be determined from instrinsic physics. In later work we will show how this introduces further freedom in the formulation of dynamical similarity, and show that there are classes of Lagrangians which are intrinsically indistinguishable. That is, to an observer subject to their dynamics and only given access to relational variables, there are many possible choices of couplings that give rise to the same observable physics. 

\section*{Acknowledgements}

\noindent The author is grateful to Julian Barbour, Jos\'e Cari\~nena, Sean Gryb and Tim Koswlowski for a number of useful discussions and comments in the preparation of this work. 

\bibliographystyle{apsrev4-1}
\bibliography{DynSimA}

\begin{thebibliography}{32}%
\makeatletter
\providecommand \@ifxundefined [1]{%
 \@ifx{#1\undefined}
}%
\providecommand \@ifnum [1]{%
 \ifnum #1\expandafter \@firstoftwo
 \else \expandafter \@secondoftwo
 \fi
}%
\providecommand \@ifx [1]{%
 \ifx #1\expandafter \@firstoftwo
 \else \expandafter \@secondoftwo
 \fi
}%
\providecommand \natexlab [1]{#1}%
\providecommand \enquote  [1]{``#1''}%
\providecommand \bibnamefont  [1]{#1}%
\providecommand \bibfnamefont [1]{#1}%
\providecommand \citenamefont [1]{#1}%
\providecommand \href@noop [0]{\@secondoftwo}%
\providecommand \href [0]{\begingroup \@sanitize@url \@href}%
\providecommand \@href[1]{\@@startlink{#1}\@@href}%
\providecommand \@@href[1]{\endgroup#1\@@endlink}%
\providecommand \@sanitize@url [0]{\catcode `\\12\catcode `\$12\catcode
  `\&12\catcode `\#12\catcode `\^12\catcode `\_12\catcode `\%12\relax}%
\providecommand \@@startlink[1]{}%
\providecommand \@@endlink[0]{}%
\providecommand \url  [0]{\begingroup\@sanitize@url \@url }%
\providecommand \@url [1]{\endgroup\@href {#1}{\urlprefix }}%
\providecommand \urlprefix  [0]{URL }%
\providecommand \Eprint [0]{\href }%
\providecommand \doibase [0]{http://dx.doi.org/}%
\providecommand \selectlanguage [0]{\@gobble}%
\providecommand \bibinfo  [0]{\@secondoftwo}%
\providecommand \bibfield  [0]{\@secondoftwo}%
\providecommand \translation [1]{[#1]}%
\providecommand \BibitemOpen [0]{}%
\providecommand \bibitemStop [0]{}%
\providecommand \bibitemNoStop [0]{.\EOS\space}%
\providecommand \EOS [0]{\spacefactor3000\relax}%
\providecommand \BibitemShut  [1]{\csname bibitem#1\endcsname}%
\let\auto@bib@innerbib\@empty
\bibitem [{\citenamefont {Mercati}(2014)}]{Mercati:2014ama}%
  \BibitemOpen
  \bibfield  {author} {\bibinfo {author} {\bibfnamefont {F.}~\bibnamefont
  {Mercati}},\ }\href@noop {} {\  (\bibinfo {year} {2014})},\ \Eprint
  {http://arxiv.org/abs/1409.0105} {arXiv:1409.0105 [gr-qc]} \BibitemShut
  {NoStop}%
\bibitem [{\citenamefont {Gomes}\ and\ \citenamefont
  {Koslowski}(2013)}]{Gomes:2012hq}%
  \BibitemOpen
  \bibfield  {author} {\bibinfo {author} {\bibfnamefont {H.}~\bibnamefont
  {Gomes}}\ and\ \bibinfo {author} {\bibfnamefont {T.}~\bibnamefont
  {Koslowski}},\ }\href {\doibase 10.1007/s10701-013-9754-0} {\bibfield
  {journal} {\bibinfo  {journal} {Found. Phys.}\ }\textbf {\bibinfo {volume}
  {43}},\ \bibinfo {pages} {1428} (\bibinfo {year} {2013})},\ \Eprint
  {http://arxiv.org/abs/1211.5878} {arXiv:1211.5878 [gr-qc]} \BibitemShut
  {NoStop}%
\bibitem [{\citenamefont {Wetterich}(2014)}]{Wetterich:2014zta}%
  \BibitemOpen
  \bibfield  {author} {\bibinfo {author} {\bibfnamefont {C.}~\bibnamefont
  {Wetterich}},\ }\href {\doibase 10.1103/PhysRevD.90.043520} {\bibfield
  {journal} {\bibinfo  {journal} {Phys. Rev.}\ }\textbf {\bibinfo {volume}
  {D90}},\ \bibinfo {pages} {043520} (\bibinfo {year} {2014})},\ \Eprint
  {http://arxiv.org/abs/1404.0535} {arXiv:1404.0535 [gr-qc]} \BibitemShut
  {NoStop}%
\bibitem [{\citenamefont {Koslowski}\ \emph {et~al.}(2018)\citenamefont
  {Koslowski}, \citenamefont {Mercati},\ and\ \citenamefont
  {Sloan}}]{Koslowski:2016hds}%
  \BibitemOpen
  \bibfield  {author} {\bibinfo {author} {\bibfnamefont {T.~A.}\ \bibnamefont
  {Koslowski}}, \bibinfo {author} {\bibfnamefont {F.}~\bibnamefont {Mercati}},
  \ and\ \bibinfo {author} {\bibfnamefont {D.}~\bibnamefont {Sloan}},\ }\href
  {\doibase 10.1016/j.physletb.2018.01.055} {\bibfield  {journal} {\bibinfo
  {journal} {Phys. Lett.}\ }\textbf {\bibinfo {volume} {B778}},\ \bibinfo
  {pages} {339} (\bibinfo {year} {2018})},\ \Eprint
  {http://arxiv.org/abs/1607.02460} {arXiv:1607.02460 [gr-qc]} \BibitemShut
  {NoStop}%
\bibitem [{\citenamefont {Cari\~nena}\ \emph {et~al.}(2013)\citenamefont
  {Cari\~nena}, \citenamefont {Falceto},\ and\ \citenamefont
  {Ra\~nada}}]{Carinena:2013zpa}%
  \BibitemOpen
  \bibfield  {author} {\bibinfo {author} {\bibfnamefont {J.~F.}\ \bibnamefont
  {Cari\~nena}}, \bibinfo {author} {\bibfnamefont {F.}~\bibnamefont {Falceto}},
  \ and\ \bibinfo {author} {\bibfnamefont {M.~F.}\ \bibnamefont {Ra\~nada}},\
  }\href@noop {} {\  (\bibinfo {year} {2013})},\ \Eprint
  {http://arxiv.org/abs/1303.6225} {arXiv:1303.6225 [math-ph]} \BibitemShut
  {NoStop}%
\bibitem [{\citenamefont {Cari\~nena}\ \emph {et~al.}(2014)\citenamefont
  {Cari\~nena}, \citenamefont {Gheorghiu}, \citenamefont {Mart\'inez},\ and\
  \citenamefont {Santos}}]{Carinena:2014bda}%
  \BibitemOpen
  \bibfield  {author} {\bibinfo {author} {\bibfnamefont {J.~F.}\ \bibnamefont
  {Cari\~nena}}, \bibinfo {author} {\bibfnamefont {I.}~\bibnamefont
  {Gheorghiu}}, \bibinfo {author} {\bibfnamefont {E.}~\bibnamefont
  {Mart\'inez}}, \ and\ \bibinfo {author} {\bibfnamefont {P.}~\bibnamefont
  {Santos}},\ }\href {\doibase 10.1088/1751-8113/47/46/465206} {\bibfield
  {journal} {\bibinfo  {journal} {J. Phys.}\ }\textbf {\bibinfo {volume}
  {A47}},\ \bibinfo {pages} {465206} (\bibinfo {year} {2014})},\ \Eprint
  {http://arxiv.org/abs/1410.2032} {arXiv:1410.2032 [math-ph]} \BibitemShut
  {NoStop}%
\bibitem [{\citenamefont {Geiges}(1997)}]{GEIGES19971193}%
  \BibitemOpen
  \bibfield  {author} {\bibinfo {author} {\bibfnamefont {H.}~\bibnamefont
  {Geiges}},\ }\href {\doibase https://doi.org/10.1016/S0040-9383(97)00004-9}
  {\bibfield  {journal} {\bibinfo  {journal} {Topology}\ }\textbf {\bibinfo
  {volume} {36}},\ \bibinfo {pages} {1193 } (\bibinfo {year}
  {1997})}\BibitemShut {NoStop}%
\bibitem [{\citenamefont {{Cari{\~n}ena}}\ \emph {et~al.}(2012)\citenamefont
  {{Cari{\~n}ena}}, \citenamefont {{Falceto}},\ and\ \citenamefont
  {{Ra{\~n}ada}}}]{2012JPhAC}%
  \BibitemOpen
  \bibfield  {author} {\bibinfo {author} {\bibfnamefont {J.~F.}\ \bibnamefont
  {{Cari{\~n}ena}}}, \bibinfo {author} {\bibfnamefont {F.}~\bibnamefont
  {{Falceto}}}, \ and\ \bibinfo {author} {\bibfnamefont {M.~F.}\ \bibnamefont
  {{Ra{\~n}ada}}},\ }\href {\doibase 10.1088/1751-8113/45/39/395210} {\bibfield
   {journal} {\bibinfo  {journal} {Journal of Physics A Mathematical General}\
  }\textbf {\bibinfo {volume} {45}},\ \bibinfo {eid} {395210} (\bibinfo {year}
  {2012})},\ \Eprint {http://arxiv.org/abs/1209.4584} {arXiv:1209.4584
  [math-ph]} \BibitemShut {NoStop}%
\bibitem [{\citenamefont {{Arnold}}\ and\ \citenamefont
  {{Novikov}}(2001)}]{ArnoldBook}%
  \BibitemOpen
  \bibfield  {author} {\bibinfo {author} {\bibfnamefont {V.~I.}\ \bibnamefont
  {{Arnold}}}\ and\ \bibinfo {author} {\bibfnamefont {S.~P.}\ \bibnamefont
  {{Novikov}}},\ }\href@noop {} {\emph {\bibinfo {title} {Dynamical systems
  IV.~Symplectic geometry and its applications, 2nd, expanded and revised
  ed.~Berlin, New York: Springer, 2001, 335 p.~Encyclopaedia of mathematical
  sciences, vol.~4, ISBN 3540626352.~Original Russian edition published by
  VINITI, Moscow, 1985}}}\ (\bibinfo  {publisher} {VINITI},\ \bibinfo {year}
  {2001})\BibitemShut {NoStop}%
\bibitem [{\citenamefont {{Etnyre}}(2001)}]{ContactIntro}%
  \BibitemOpen
  \bibfield  {author} {\bibinfo {author} {\bibfnamefont {J.~B.}\ \bibnamefont
  {{Etnyre}}},\ }\href@noop {} {\  (\bibinfo {year} {2001})},\ \Eprint
  {http://arxiv.org/abs/0111118} {arXiv:0111118 [math]} \BibitemShut {NoStop}%
\bibitem [{\citenamefont {Geiges}(2008)}]{geiges2008introduction}%
  \BibitemOpen
  \bibfield  {author} {\bibinfo {author} {\bibfnamefont {H.}~\bibnamefont
  {Geiges}},\ }\href@noop {} {\emph {\bibinfo {title} {An introduction to
  contact topology}}},\ Vol.\ \bibinfo {volume} {109}\ (\bibinfo  {publisher}
  {Cambridge University Press},\ \bibinfo {year} {2008})\BibitemShut {NoStop}%
\bibitem [{\citenamefont {Bravetti}\ \emph {et~al.}(2017)\citenamefont
  {Bravetti}, \citenamefont {Cruz},\ and\ \citenamefont {Tapias}}]{Contact1}%
  \BibitemOpen
  \bibfield  {author} {\bibinfo {author} {\bibfnamefont {A.}~\bibnamefont
  {Bravetti}}, \bibinfo {author} {\bibfnamefont {H.}~\bibnamefont {Cruz}}, \
  and\ \bibinfo {author} {\bibfnamefont {D.}~\bibnamefont {Tapias}},\ }\href
  {\doibase 10.1016/j.aop.2016.11.003} {\bibfield  {journal} {\bibinfo
  {journal} {Annals of Physics}\ }\textbf {\bibinfo {volume} {376}},\ \bibinfo
  {pages} {17} (\bibinfo {year} {2017})},\ \Eprint
  {http://arxiv.org/abs/1604.08266} {arXiv:1604.08266 [math-ph]} \BibitemShut
  {NoStop}%
\bibitem [{\citenamefont {{Bravetti}}\ and\ \citenamefont
  {{Tapias}}(2015)}]{ContactLiouville}%
  \BibitemOpen
  \bibfield  {author} {\bibinfo {author} {\bibfnamefont {A.}~\bibnamefont
  {{Bravetti}}}\ and\ \bibinfo {author} {\bibfnamefont {D.}~\bibnamefont
  {{Tapias}}},\ }\href {\doibase 10.1088/1751-8113/48/24/245001} {\bibfield
  {journal} {\bibinfo  {journal} {Journal of Physics A Mathematical General}\
  }\textbf {\bibinfo {volume} {48}},\ \bibinfo {eid} {245001} (\bibinfo {year}
  {2015})},\ \Eprint {http://arxiv.org/abs/1412.0026} {arXiv:1412.0026
  [math-ph]} \BibitemShut {NoStop}%
\bibitem [{\citenamefont {Barbour}\ \emph {et~al.}(2014)\citenamefont
  {Barbour}, \citenamefont {Koslowski},\ and\ \citenamefont
  {Mercati}}]{Barbour:2014bga}%
  \BibitemOpen
  \bibfield  {author} {\bibinfo {author} {\bibfnamefont {J.}~\bibnamefont
  {Barbour}}, \bibinfo {author} {\bibfnamefont {T.}~\bibnamefont {Koslowski}},
  \ and\ \bibinfo {author} {\bibfnamefont {F.}~\bibnamefont {Mercati}},\ }\href
  {\doibase 10.1103/PhysRevLett.113.181101} {\bibfield  {journal} {\bibinfo
  {journal} {Phys. Rev. Lett.}\ }\textbf {\bibinfo {volume} {113}},\ \bibinfo
  {pages} {181101} (\bibinfo {year} {2014})},\ \Eprint
  {http://arxiv.org/abs/1409.0917} {arXiv:1409.0917 [gr-qc]} \BibitemShut
  {NoStop}%
\bibitem [{\citenamefont {Barbour}\ \emph {et~al.}(2013)\citenamefont
  {Barbour}, \citenamefont {Koslowski},\ and\ \citenamefont
  {Mercati}}]{Barbour:2013jya}%
  \BibitemOpen
  \bibfield  {author} {\bibinfo {author} {\bibfnamefont {J.}~\bibnamefont
  {Barbour}}, \bibinfo {author} {\bibfnamefont {T.}~\bibnamefont {Koslowski}},
  \ and\ \bibinfo {author} {\bibfnamefont {F.}~\bibnamefont {Mercati}},\
  }\href@noop {} {\  (\bibinfo {year} {2013})},\ \Eprint
  {http://arxiv.org/abs/1310.5167} {arXiv:1310.5167 [gr-qc]} \BibitemShut
  {NoStop}%
\bibitem [{\citenamefont {Gomes}\ \emph {et~al.}(2011)\citenamefont {Gomes},
  \citenamefont {Gryb},\ and\ \citenamefont {Koslowski}}]{Gomes:2010fh}%
  \BibitemOpen
  \bibfield  {author} {\bibinfo {author} {\bibfnamefont {H.}~\bibnamefont
  {Gomes}}, \bibinfo {author} {\bibfnamefont {S.}~\bibnamefont {Gryb}}, \ and\
  \bibinfo {author} {\bibfnamefont {T.}~\bibnamefont {Koslowski}},\ }\href
  {\doibase 10.1088/0264-9381/28/4/045005} {\bibfield  {journal} {\bibinfo
  {journal} {Class. Quant. Grav.}\ }\textbf {\bibinfo {volume} {28}},\ \bibinfo
  {pages} {045005} (\bibinfo {year} {2011})},\ \Eprint
  {http://arxiv.org/abs/1010.2481} {arXiv:1010.2481 [gr-qc]} \BibitemShut
  {NoStop}%
\bibitem [{\citenamefont {Gryb}\ and\ \citenamefont
  {Thebault}(2016)}]{Gryb:2014mva}%
  \BibitemOpen
  \bibfield  {author} {\bibinfo {author} {\bibfnamefont {S.}~\bibnamefont
  {Gryb}}\ and\ \bibinfo {author} {\bibfnamefont {K.}~\bibnamefont
  {Thebault}},\ }\href {\doibase 10.1093/bjps/axv009} {\bibfield  {journal}
  {\bibinfo  {journal} {Brit. J. Phil. Sci.}\ }\textbf {\bibinfo {volume}
  {67}},\ \bibinfo {pages} {663} (\bibinfo {year} {2016})},\ \Eprint
  {http://arxiv.org/abs/1408.2691} {arXiv:1408.2691 [gr-qc]} \BibitemShut
  {NoStop}%
\bibitem [{\citenamefont {Gomes}(2016)}]{Gomes:2016aam}%
  \BibitemOpen
  \bibfield  {author} {\bibinfo {author} {\bibfnamefont {H.~d.~A.}\
  \bibnamefont {Gomes}},\ }\href@noop {} {\  (\bibinfo {year} {2016})},\
  \Eprint {http://arxiv.org/abs/1603.01574} {arXiv:1603.01574 [quant-ph]}
  \BibitemShut {NoStop}%
\bibitem [{\citenamefont {Evans}\ \emph {et~al.}(2016)\citenamefont {Evans},
  \citenamefont {Gryb},\ and\ \citenamefont {Th\'ebault}}]{Evans:2016org}%
  \BibitemOpen
  \bibfield  {author} {\bibinfo {author} {\bibfnamefont {P.~W.}\ \bibnamefont
  {Evans}}, \bibinfo {author} {\bibfnamefont {S.}~\bibnamefont {Gryb}}, \ and\
  \bibinfo {author} {\bibfnamefont {K.~P.~Y.}\ \bibnamefont {Th\'ebault}},\
  }\href {\doibase 10.1016/j.shpsb.2016.10.005} {\bibfield  {journal} {\bibinfo
   {journal} {Stud. Hist. Phil. Sci.}\ }\textbf {\bibinfo {volume} {B56}},\
  \bibinfo {pages} {1} (\bibinfo {year} {2016})},\ \Eprint
  {http://arxiv.org/abs/1606.07265} {arXiv:1606.07265 [gr-qc]} \BibitemShut
  {NoStop}%
\bibitem [{\citenamefont {Earman}(2006)}]{EARMAN2006399}%
  \BibitemOpen
  \bibfield  {author} {\bibinfo {author} {\bibfnamefont {J.}~\bibnamefont
  {Earman}},\ }\href {\doibase https://doi.org/10.1016/j.shpsb.2006.03.002}
  {\bibfield  {journal} {\bibinfo  {journal} {Studies in History and Philosophy
  of Science Part B: Studies in History and Philosophy of Modern Physics}\
  }\textbf {\bibinfo {volume} {37}},\ \bibinfo {pages} {399 } (\bibinfo {year}
  {2006})},\ \bibinfo {note} {the arrows of time, 2006}\BibitemShut {NoStop}%
\bibitem [{\citenamefont {Corichi}\ and\ \citenamefont
  {Sloan}(2014)}]{Corichi:2013kua}%
  \BibitemOpen
  \bibfield  {author} {\bibinfo {author} {\bibfnamefont {A.}~\bibnamefont
  {Corichi}}\ and\ \bibinfo {author} {\bibfnamefont {D.}~\bibnamefont
  {Sloan}},\ }\href {\doibase 10.1088/0264-9381/31/6/062001} {\bibfield
  {journal} {\bibinfo  {journal} {Class. Quant. Grav.}\ }\textbf {\bibinfo
  {volume} {31}},\ \bibinfo {pages} {062001} (\bibinfo {year} {2014})},\
  \Eprint {http://arxiv.org/abs/1310.6399} {arXiv:1310.6399 [gr-qc]}
  \BibitemShut {NoStop}%
\bibitem [{\citenamefont {Sloan}(2014)}]{Sloan:2014jra}%
  \BibitemOpen
  \bibfield  {author} {\bibinfo {author} {\bibfnamefont {D.}~\bibnamefont
  {Sloan}},\ }\href {\doibase 10.1088/0264-9381/31/24/245015} {\bibfield
  {journal} {\bibinfo  {journal} {Class. Quant. Grav.}\ }\textbf {\bibinfo
  {volume} {31}},\ \bibinfo {pages} {245015} (\bibinfo {year} {2014})},\
  \Eprint {http://arxiv.org/abs/1407.3977} {arXiv:1407.3977 [gr-qc]}
  \BibitemShut {NoStop}%
\bibitem [{\citenamefont {Sloan}(2016)}]{Sloan:2016nnx}%
  \BibitemOpen
  \bibfield  {author} {\bibinfo {author} {\bibfnamefont {D.}~\bibnamefont
  {Sloan}},\ }\href {\doibase 10.21105/astro.1602.02113} {\  (\bibinfo {year}
  {2016}),\ 10.21105/astro.1602.02113},\ \Eprint
  {http://arxiv.org/abs/1602.02113} {arXiv:1602.02113 [gr-qc]} \BibitemShut
  {NoStop}%
\bibitem [{\citenamefont {Ashtekar}(2013)}]{Ashtekar:2012np}%
  \BibitemOpen
  \bibfield  {author} {\bibinfo {author} {\bibfnamefont {A.}~\bibnamefont
  {Ashtekar}},\ }\bibfield  {booktitle} {\emph {\bibinfo {booktitle}
  {{Proceedings, 3rd Quantum Geometry and Quantum Gravity School: Zakopane,
  Poland, February 28-March 13, 2011}}},\ }\href {\doibase
  10.1007/978-3-642-33036-0_2} {\bibfield  {journal} {\bibinfo  {journal}
  {Lect. Notes Phys.}\ }\textbf {\bibinfo {volume} {863}},\ \bibinfo {pages}
  {31} (\bibinfo {year} {2013})},\ \Eprint {http://arxiv.org/abs/1201.4598}
  {arXiv:1201.4598 [gr-qc]} \BibitemShut {NoStop}%
\bibitem [{\citenamefont {Sloan}\ and\ \citenamefont
  {Silk}(2016)}]{Sloan:2015bha}%
  \BibitemOpen
  \bibfield  {author} {\bibinfo {author} {\bibfnamefont {D.}~\bibnamefont
  {Sloan}}\ and\ \bibinfo {author} {\bibfnamefont {J.}~\bibnamefont {Silk}},\
  }\href {\doibase 10.1103/PhysRevD.93.104030} {\bibfield  {journal} {\bibinfo
  {journal} {Phys. Rev.}\ }\textbf {\bibinfo {volume} {D93}},\ \bibinfo {pages}
  {104030} (\bibinfo {year} {2016})},\ \Eprint
  {http://arxiv.org/abs/1505.01445} {arXiv:1505.01445 [gr-qc]} \BibitemShut
  {NoStop}%
\bibitem [{\citenamefont {Curiel}(2015)}]{Curiel:2015oea}%
  \BibitemOpen
  \bibfield  {author} {\bibinfo {author} {\bibfnamefont {E.}~\bibnamefont
  {Curiel}},\ }\href@noop {} {\  (\bibinfo {year} {2015})},\ \Eprint
  {http://arxiv.org/abs/1509.01878} {arXiv:1509.01878 [gr-qc]} \BibitemShut
  {NoStop}%
\bibitem [{\citenamefont {Ashtekar}\ and\ \citenamefont
  {Sloan}(2010)}]{Measure}%
  \BibitemOpen
  \bibfield  {author} {\bibinfo {author} {\bibfnamefont {A.}~\bibnamefont
  {Ashtekar}}\ and\ \bibinfo {author} {\bibfnamefont {D.}~\bibnamefont
  {Sloan}},\ }\href {\doibase 10.1016/j.physletb.2010.09.058} {\bibfield
  {journal} {\bibinfo  {journal} {Phys.Lett.}\ }\textbf {\bibinfo {volume}
  {B694}},\ \bibinfo {pages} {108} (\bibinfo {year} {2010})},\ \Eprint
  {http://arxiv.org/abs/0912.4093} {arXiv:0912.4093 [gr-qc]} \BibitemShut
  {NoStop}%
\bibitem [{\citenamefont {Corichi}\ and\ \citenamefont
  {Karami}(2011)}]{Corichi:2010zp}%
  \BibitemOpen
  \bibfield  {author} {\bibinfo {author} {\bibfnamefont {A.}~\bibnamefont
  {Corichi}}\ and\ \bibinfo {author} {\bibfnamefont {A.}~\bibnamefont
  {Karami}},\ }\href {\doibase 10.1103/PhysRevD.83.104006} {\bibfield
  {journal} {\bibinfo  {journal} {Phys.Rev.}\ }\textbf {\bibinfo {volume}
  {D83}},\ \bibinfo {pages} {104006} (\bibinfo {year} {2011})},\ \Eprint
  {http://arxiv.org/abs/1011.4249} {arXiv:1011.4249 [gr-qc]} \BibitemShut
  {NoStop}%
\bibitem [{\citenamefont {Ashtekar}\ and\ \citenamefont
  {Sloan}(2011)}]{Measure2}%
  \BibitemOpen
  \bibfield  {author} {\bibinfo {author} {\bibfnamefont {A.}~\bibnamefont
  {Ashtekar}}\ and\ \bibinfo {author} {\bibfnamefont {D.}~\bibnamefont
  {Sloan}},\ }\href {\doibase 10.1007/s10714-011-1246-y} {\bibfield  {journal}
  {\bibinfo  {journal} {General Relativity and Gravitation}\ }\textbf {\bibinfo
  {volume} {43}} (\bibinfo {year} {2011}),\ 10.1007/s10714-011-1246-y},\
  \bibinfo {note}
  {\url{http://dx.doi.org/10.1007/s10714-011-1246-y}}\BibitemShut {NoStop}%
\bibitem [{\citenamefont {Bianchi}\ \emph {et~al.}(2018)\citenamefont
  {Bianchi}, \citenamefont {Christodoulou}, \citenamefont {D'Ambrosio},
  \citenamefont {Rovelli},\ and\ \citenamefont {Haggard}}]{Bianchi:2018mml}%
  \BibitemOpen
  \bibfield  {author} {\bibinfo {author} {\bibfnamefont {E.}~\bibnamefont
  {Bianchi}}, \bibinfo {author} {\bibfnamefont {M.}~\bibnamefont
  {Christodoulou}}, \bibinfo {author} {\bibfnamefont {F.}~\bibnamefont
  {D'Ambrosio}}, \bibinfo {author} {\bibfnamefont {C.}~\bibnamefont {Rovelli}},
  \ and\ \bibinfo {author} {\bibfnamefont {H.~M.}\ \bibnamefont {Haggard}},\
  }\href@noop {} {\  (\bibinfo {year} {2018})},\ \Eprint
  {http://arxiv.org/abs/1802.04264} {arXiv:1802.04264 [gr-qc]} \BibitemShut
  {NoStop}%
\bibitem [{\citenamefont {D'Ambrosio}\ and\ \citenamefont
  {Rovelli}(tion)}]{CarloPrep}%
  \BibitemOpen
  \bibfield  {author} {\bibinfo {author} {\bibfnamefont {F.}~\bibnamefont
  {D'Ambrosio}}\ and\ \bibinfo {author} {\bibfnamefont {C.}~\bibnamefont
  {Rovelli}},\ }\href@noop {} {\  (\bibinfo {year} {in
  preparation})}\BibitemShut {NoStop}%
\bibitem [{\citenamefont {Koslowski}(tion)}]{TimPrep}%
  \BibitemOpen
  \bibfield  {author} {\bibinfo {author} {\bibfnamefont {T.}~\bibnamefont
  {Koslowski}},\ }\href@noop {} {\  (\bibinfo {year} {in
  preparation})}\BibitemShut {NoStop}%
\end{thebibliography}%

\end{document}